\documentclass{article}

\usepackage[preprint]{neurips_2024}

\usepackage[utf8]{inputenc} 
\usepackage[T1]{fontenc}    
\usepackage{hyperref}       
\usepackage{url}            
\usepackage{booktabs}       
\usepackage{amsfonts}       
\usepackage{nicefrac}       
\usepackage{microtype}      
\usepackage{xcolor}         

\usepackage{amsmath,amsfonts,bm}

\def\eqref#1{equation~\ref{#1}}

\def\1{\bm{1}}

\DeclareMathAlphabet{\mathsfit}{\encodingdefault}{\sfdefault}{m}{sl}
\SetMathAlphabet{\mathsfit}{bold}{\encodingdefault}{\sfdefault}{bx}{n}

\newcommand{\R}{\mathbb{R}}

\newcommand{\Var}{\mathrm{Var}}

\usepackage{graphicx}
\graphicspath{{./figures/}} 
\usepackage{amsmath}
\usepackage{cleveref} 

\usepackage{algorithm}
\usepackage{algpseudocode}
\usepackage{caption}

\title{Unsupervised Evolutionary Cell Type Matching via Entropy-Minimized Optimal Transport}

\author{%
Mu Qiao \\
Meta Platforms \\
\texttt{muqiao0626@gmail.com}
}

\begin{document}

\maketitle

\begin{abstract}
Identifying evolutionary correspondences between cell types across species is a fundamental challenge in comparative genomics and evolutionary biology. Existing approaches often rely on either reference-based matching, which imposes asymmetry by designating one species as the reference, or projection-based matching, which may increase computational complexity and obscure biological interpretability at the cell-type level. Here, we present OT-MESH, an unsupervised computational framework leveraging entropy-regularized optimal transport (OT) to systematically determine cross-species cell type homologies. Our method uniquely integrates the Minimize Entropy of Sinkhorn (MESH) technique to refine the OT plan, transforming diffuse transport matrices into sparse, interpretable correspondences. Through systematic evaluation on synthetic datasets, we demonstrate that OT-MESH achieves near-optimal matching accuracy with computational efficiency, while maintaining remarkable robustness to noise. Compared to other OT-based methods like RefCM, OT-MESH provides speedup while achieving comparable accuracy. Applied to retinal bipolar cells (BCs) and retinal ganglion cells (RGCs) from mouse and macaque, OT-MESH accurately recovers known evolutionary relationships and uncovers novel correspondences, one of which was independently validated experimentally. Thus, our framework offers a principled, scalable, and interpretable solution for evolutionary cell type mapping, facilitating deeper insights into cellular specialization and conservation across species.
\end{abstract}

\section{Introduction}
Understanding evolutionary relationships between cell types across species is central to elucidating mechanisms underlying cellular specialization and diversity. Recent advances in single-cell RNA sequencing (scRNA-seq) have enabled unprecedented characterization of cell types based on their gene expression profiles across diverse species~\cite{wangTracingCelltypeEvolution2021,hahnEvolutionNeuronalCell2023a}, with atlas-scale datasets now containing hundreds to thousands of distinct cell types. However, a fundamental challenge persists: given independently identified clusters of transcriptionally similar cells from different species, how can we systematically, reliably, and efficiently determine their evolutionary correspondences?

Existing methods for cross-species cell type alignment broadly fall into two categories~\cite{shaferCrossSpeciesAnalysisSingleCell2019}: (1) reference-based matching, in which classifiers are trained on labeled data from a reference species to predict cell types in a query species~\cite{shekharCOMPREHENSIVECLASSIFICATIONRETINAL2016a,pengMolecularClassificationComparative2019a,zhangEvolutionaryDevelopmentalSpecialization2024,wangMolecularCharacterizationSea2024}; and (2) projection-based matching, in which individual-cell gene expression profiles from different species are embedded into a shared low-dimensional space using algorithms such as Canonical Correlation Analysis (CCA), Harmony, or other manifold alignment techniques~\cite{butlerIntegratingSinglecellTranscriptomic2018b,korsunskyFastSensitiveAccurate2019,rosenUniversalCellEmbeddings2024a}. 

Reference-based matching inherently introduces asymmetry by designating one species as the "reference" and the other as the "query," potentially biasing outcomes, especially when species differ significantly in cell type numbers or evolutionary distances. Projection-based matching, though symmetric, typically operates at the level of individual cells, resulting in high computational demands and potentially obscuring direct biological interpretability of cell-type level relationships. Recent optimal transport (OT) based methods like RefCM~\cite{galantiAutomatedCellType2024} offer improved theoretical foundations but rely on computationally expensive exact earth mover's distance (EMD) calculations, limiting their scalability for large atlases.

To address these limitations, we introduce OT-MESH, an unsupervised computational framework that combines entropy-regularized OT with the Minimize Entropy of Sinkhorn (MESH) refinement strategy~\cite{zhangUnlockingSlotAttention2023}. The key insight is that while standard entropy-regularized OT is computationally efficient but produces diffuse, non-sparse transport plans unsuitable for biological interpretation. The MESH technique iteratively refines the cost matrix to promote sparsity in the cross-species cell type alignment while maintaining the computational advantages of entropy regularization. This approach enables OT-MESH to achieve the interpretability while maintaining the speed necessary for large-scale comparative studies.

We illustrate our approach through comprehensive experiments on both synthetic and biological datasets. Using synthetic data with known ground truth correspondences, we demonstrate that OT-MESH maintains high accuracy with sub-second runtime for up to 500 cell types—54× faster than RefCM while achieving comparable performance. The method shows remarkable robustness, maintaining high matching accuracy even at 8× noise levels, substantially outperforming projection-based and supervised approaches. Applied to established datasets of bipolar cells (BCs) and retinal ganglion cells (RGCs) from mouse and macaque retinas~\cite{hahnEvolutionNeuronalCell2023a,pengMolecularClassificationComparative2019a,tranSinglecellProfilesRetinal2019a}, OT-MESH consistently identifies known evolutionary correspondences and discovers novel cell-type homologies, including one correspondence independently validated experimentally~\cite{wangONtypeDirectionselectiveGanglion2023}, demonstrating the method's biological relevance. Thus, OT-MESH provides a principled, scalable, and interpretable computational framework, enhancing our ability to map cell type evolution and advance its understanding.

\section{Background}
\subsection{Optimal Transport (OT) for Correspondence Problems}
OT provides a principled mathematical framework for identifying correspondence between distributions. Given two distributions supported on spaces \(\mathbf{X}\) and \(\mathbf{Y}\), OT seeks the optimal mapping that minimizes the total cost of transporting mass from \(\mathbf{X}\) to \(\mathbf{Y}\). In discrete form, it involves finding a transport plan \(\mathbf{W}\) that minimizes 
\begin{equation}
\label{eq:OT}
L(\mathbf{W}) = \sum_{ij} W_{ij}C_{ij},
\end{equation}
where \(C_{ij}\) is the cost of transporting unit mass from point \(i\) in \(\mathbf{X}\) to point \(j\) in \(\mathbf{Y}\).

To enhance computational tractability and numerical stability, OT problems often incorporate entropy regularization, which encourages smoother transport plans. The entropy-regularized OT problem is formulated as:
\begin{equation}
\label{eq:OT_reg}
L(\mathbf{W}) = \sum_{ij} W_{ij}C_{ij} + \alpha\sum_{ij} W_{ij} \Bigl(\log W_{ij} - 1\Bigr),
\end{equation}
subject to the marginal constraints:
\begin{equation}
\label{eq:marginalConstraints}
\sum_{j}W_{ij} = p_i,\quad \sum_{i}W_{ij} = q_j.
\end{equation}
Here, $p_i$ and $q_j$ define prescribed marginals. This entropy-regularized problem can be efficiently solved using the Sinkhorn algorithm~\cite{cuturiSinkhornDistancesLightspeed2013}.

\subsubsection{Sinkhorn Algorithm}
Given a cost matrix \(\mathbf{C}\in\R^{a\times b}\), a regularization parameter \(\alpha>0\), and prescribed marginal distributions \(\mathbf{p}\in\R^a\) and \(\mathbf{q}\in\R^b\), the Sinkhorn algorithm computes the Gibbs kernel as: 
\begin{equation}
\label{eq:GibbsKernel}
\mathbf{G} = \exp\left(-\frac{\mathbf{C}}{\alpha}\right).
\end{equation}
It iteratively updates two scaling vectors \(\mathbf{u}\) and \(\mathbf{v}\) as follows:
\begin{equation}
\label{eq:scalingVectors}
\mathbf{u} \leftarrow \mathbf{p} \oslash (\mathbf{G}\mathbf{v}), \qquad \mathbf{v} \leftarrow \mathbf{q} \oslash (\mathbf{G}^\top \mathbf{u}),
\end{equation}
where \(\oslash\) denotes element-wise division. After convergence, the final transport matrix is obtained by:
\begin{equation}
\label{eq:finalSinkhorn}
\mathbf{W} = \operatorname{diag}(\mathbf{u})\,\mathbf{G}\,\operatorname{diag}(\mathbf{v}).
\end{equation}
In practice, computations are frequently performed in log-space to maintain numerical stability.

\subsection{Minimize Entropy of Sinkhorn (MESH)}
While entropy regularization makes OT computationally tractable, it often produces overly diffuse (low interpretability) transport plans with limited interpretability, especially when the underlying biological correspondences are expected to be sparse. To address this, the MESH approach~\cite{zhangUnlockingSlotAttention2023} iteratively refines the cost matrix itself to minimize the entropy of the resulting transport plan:
\begin{equation}
\label{eq:MESHUpdateCost}
\mathbf{C}'(t+1) = \mathbf{C}'(t) - \lambda \frac{\nabla_{\mathbf{C}'(t)} H\bigl(\text{sinkhorn}(\mathbf{C}'(t))\bigr)}{\|\nabla_{\mathbf{C}'(t)} H\bigl(\text{sinkhorn}(\mathbf{C}'(t))\bigr)\|},
\end{equation}
where the entropy of the transport plan $\mathbf{W}(t) = \text{sinkhorn}(\mathbf{C}'(t))$ is defined as:
\begin{equation}
\label{eq:entropyOfPlan}
H(\mathbf{W}(t)) = - \sum_{ij} W_{ij}(t)\log W_{ij}(t),
\end{equation}
and \(\lambda\) is a learning rate. This iterative approach yields sparse and interpretable correspondence matrices by concentrating mass onto fewer, more meaningful connections.

\subsection{OT Applications in Single-Cell Biology}
OT methods have found increasing applications in single-cell biology, particularly for aligning cell states across experimental conditions, measurement modalities, or developmental time points~\cite{demetciSCOTSingleCellMultiOmics2022,gossiMatchingSingleCells2023,galantiAutomatedCellType2024,penaConstructingCelltypeTaxonomy2025}. A key advantage of OT approaches lies in their natural ability to handle partial or incomplete correspondences, and recent studies have leveraged OT to construct hierarchical cell-type taxonomies or annotations~\cite{galantiAutomatedCellType2024,penaConstructingCelltypeTaxonomy2025}. Specifically, Galanti et al.~\cite{galantiAutomatedCellType2024} introduced reference cluster mapping (RefCM), which combines optimal transport with integer programming to enforce sparse mappings between reference and query datasets. This approach addresses the over-diffusion problem in standard OT by solving a secondary discrete optimization problem, allowing for complex relationships including many-to-one and one-to-many mappings between cell types. While RefCM and other OT-based methods have shown promise for within-species cell type annotation, they have yet to be comprehensively validated specifically for cross-species evolutionary cell type comparisons. In this study, we address this gap by adapting and extending OT methodologies, particularly through the MESH procedure, to the problem of cross-species cell type matching.

\section{Methods}

\subsection{Problem Formulation}
We consider two species containing \(a\) and \(b\) distinct cell types, indexed by \(i\) and \(j\) respectively. For the \(i\)th cell type in the first species, let \(\mathbf{x}_{ik} \in \R^{1 \times g}\) represent the gene expression vector of the the \(k\)th cell; similarly, let \(\mathbf{y}_{jl} \in \R^{1 \times g}\) denote expression for the \(l\)th cell of the \(j\)th type in the second species. Our goal is to identify an optimal correspondence matrix \(\mathbf{W} \in \R^{a \times b}\) where each element \(W_{ij}\) assigns a probability to each pair of cell types being evolutionarily related.

\subsection{Gene Selection by Signal-to-Noise Ratio (SNR)}
Rather than using all available genes, which can introduce noise and increase computational burden, we identify a subset of highly informative genes based on their expression SNR~\cite{hahnEvolutionNeuronalCell2023a}. SNR prioritizes genes that show high variance between cell types relative to variance within cell types, thus highlighting genes that effectively distinguish types. Specifically, for each gene \(g\), intra-type variances \(s^{2}_{i,g}\) and \(s^{2}_{j,g}\) are computed as: 
\begin{equation}
\label{eq:varianceSpecies1}
s^{2}_{i,g} = \frac{\sum_{k}\left(x_{ik,g}-\bar{x}_{i:,g}\right)^2}{n_i}, \quad s^{2}_{j,g} = \frac{\sum_{l}\left(y_{jl,g}-\bar{y}_{j:,g}\right)^2}{n_j}.
\end{equation}
where \(\bar{x}_{i:,g}\) and \(\bar{y}_{j:,g}\) denote the mean expression level of gene \(g\) for cell types $i$ and $j$ respectively, and $n_i$, $n_j$ represent the numbers of cells in those types. The SNR for each gene \(g\) is calculated by:
\begin{equation}
\label{eq:SNR}
\text{SNR}(g) = \frac{\Var\left(\{\bar{x}_{1:,g},\dots,\bar{x}_{a:,g},\bar{y}_{1:,g},\dots,\bar{y}_{b:,g}\}\right)}{\frac{1}{a+b}\left(\sum_{i=1}^{a}{s^{2}_{i,g}}+\sum_{j=1}^{b}{s^{2}_{j,g}}\right)}.
\end{equation}
We retain genes exceeding a predefined SNR threshold, resulting in $m$ selected genes used to construct the centroid matrices \(\hat{\mathbf{X}} \in \R^{a \times m}\) for the first species, and \(\hat{\mathbf{Y}} \in \R^{b \times m}\) for the second species.

\subsection{Cost Matrix Construction}
We define a cost matrix \(\mathbf{C}\in\R^{a \times b}\) to quantify the dissimilarity between cell types across species. This cost is based on the cosine distance between the high-dimensional gene expression centroids:
\begin{equation}
\label{eq:costMatrix}
C_{ij} = 1 - \frac{\hat{\mathbf{X}}_{i} \cdot \hat{\mathbf{Y}}_{j}}{\|\hat{\mathbf{X}}_{i}\|\,\|\hat{\mathbf{Y}}_{j}\|}.
\end{equation}
Here, \(\hat{\mathbf{X}}_{i}\) and \(\hat{\mathbf{Y}}_{j}\) are the centroid vectors of cell types \(i\) and \(j\) respectively.

Our choice of cosine distance is deliberate. In cross-species comparisons, homologous cell types may preserve the relative pattern of their gene expression profiles but exhibit significant differences in the absolute magnitude of expression due to technical variability (e.g., sequencing depth) or biological divergence. Cosine distance is critically insensitive to these magnitude shifts, as it only measures the angle between two vectors. In contrast, a metric like Euclidean distance would be confounded by such scaling effects, potentially assigning a high dissimilarity cost to genuinely related cell types. Thus, cosine distance provides a robust measure of profile similarity.

\subsection{Entropy-Regularized OT with MESH}
To obtain sparse and interpretable correspondences, we employ entropy-regularized OT refined by the MESH procedure. We begin with a slightly perturbed initial cost matrix~\cite{zhangUnlockingSlotAttention2023}:
\begin{equation}
\label{eq:meshInit}
\mathbf{C}'(0) = \mathbf{C} + \varepsilon,\quad \varepsilon_{ij} \sim \mathcal{N}(0, 10^{-6}),
\end{equation}

At each iteration \(t\), we compute the transport plan \(\mathbf{W}(t)\) from the current cost matrix \(\mathbf{C}'(t)\) using the Sinkhorn algorithm and then quantify its entropy:
\begin{equation}
\label{eq:entropyFunction} 
F\bigl(\mathbf{C}'(t)\bigr) = H(\mathbf{W}(t)) = -\sum_{ij}W_{ij}(t) \log W_{ij}(t).
\end{equation}
The cost matrix is updated by gradient descent, normalized to maintain stable updates:
\begin{equation}
\label{eq:meshUpdate} 
\mathbf{C}'(t+1) = \mathbf{C}'(t) - \lambda \frac{\nabla_{\mathbf{C}'(t)} F\bigl(\mathbf{C}'(t)\bigr)}{\|\nabla_{\mathbf{C}'(t)} F\bigl(\mathbf{C}'(t)\bigr)\|},
\end{equation}
After \(T\) iterations, the resulting cost matrix \(\mathbf{C}'(T)\) is used to compute the final transport plan:
\begin{equation}
\label{eq:finalW}
\mathbf{W}^* = \text{sinkhorn}(\mathbf{C}'(T)).
\end{equation}

\subsection{Algorithm Summary}

\begin{figure}[htbp]
\centering
\includegraphics[trim=0 250 100 0, clip, width=0.95\linewidth]{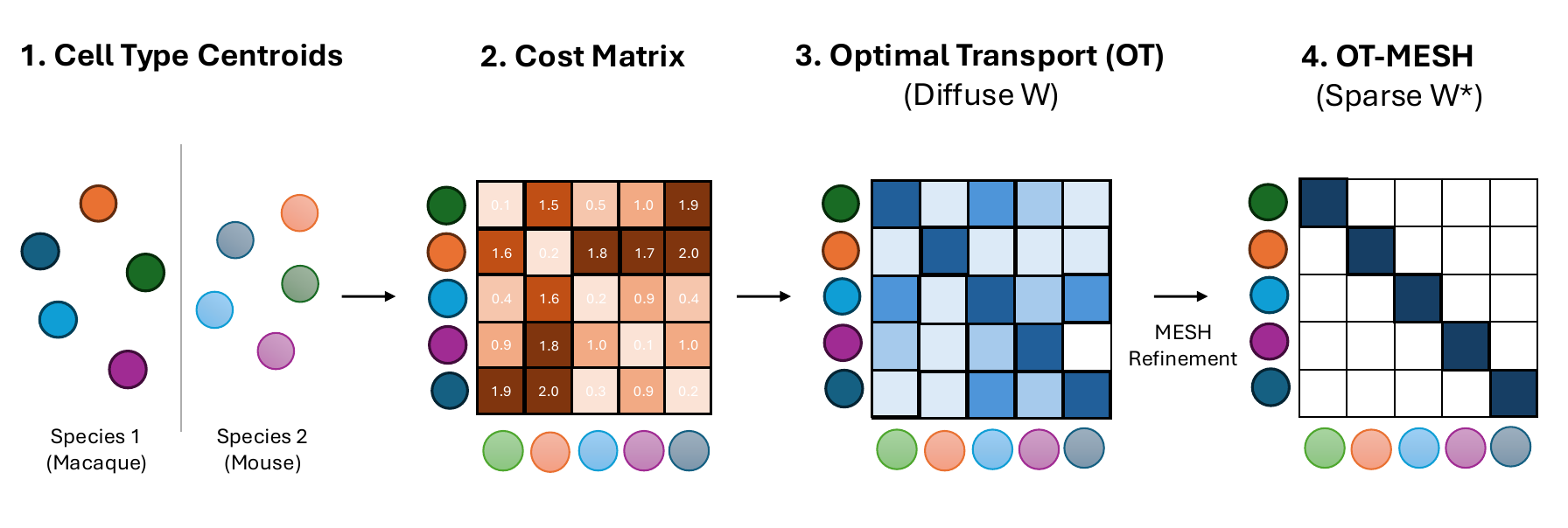} 
\caption{Overview of the OT-MESH framework for cross-species cell type matching. Starting from cell type centroids computed using SNR-selected genes, a cost matrix is constructed based on cosine distances between centroids from two species. Standard entropy-regularized OT yields an initial correspondence matrix \(\mathbf{W}\) that is typically diffuse and difficult to interpret biologically. The MESH procedure iteratively refines the cost matrix through entropy minimization, transforming the diffuse transport plan into a sparse, interpretable correspondence matrix \(\mathbf{W}^*\) that clearly identifies evolutionarily related cell types between species.}
\label{fig1}
\end{figure}

Our entire approach is summarized as Algorithm~\ref{alg:main} and illustrated schematically in Figure~\ref{fig1}.

\begin{algorithm}[htbp] 
\caption{Evolutionary Cell Type Matching via Entropy-Regularized OT with MESH}
\label{alg:main}
\begin{algorithmic}[1]
\State \textbf{Input:} Centroid matrices from SNR-selected genes, \(\hat{\mathbf{X}} \in \R^{a \times m}\) and \(\hat{\mathbf{Y}} \in \R^{b \times m}\), regularization parameter \(\alpha\), MESH learning rate \(\lambda\), number of MESH iterations \(T\)
\State Compute cost matrix \(\mathbf{C} \in \R^{a \times b}\) as in Eq.~\ref{eq:costMatrix}
\State Initialize marginal distributions \(\mathbf{p} \), \(\mathbf{q}\), and \(\mathbf{C}'(0)\) as in Eq.~\ref{eq:meshInit}
\State For \(t = 0\) to \(T-1\):
\State \hspace{\algorithmicindent} Compute \(\mathbf{W}(t) = \text{sinkhorn}(\mathbf{C}'(t))\)
\State \hspace{\algorithmicindent} Calculate entropy \(F\bigl(\mathbf{C}'(t)\bigr)\) as in Eq.~\ref{eq:entropyFunction}
\State \hspace{\algorithmicindent} Update \(\mathbf{C}'(t+1)\) using Eq.~\ref{eq:meshUpdate}
\State \textbf{Output:} Final correspondence matrix \(\mathbf{W}^*\) as in Eq.~\ref{eq:finalW}
\end{algorithmic}
\end{algorithm}

\section{Experimental Setup}

\subsection{Datasets}

\subsubsection{Synthetic Data for Scalability and Robustness Analysis}
To systematically evaluate computational scalability and noise robustness, we generated synthetic matched datasets simulating cross-species correspondence scenarios. These datasets consisted of paired "species" with known 1:1 correspondences, where expression profiles were generated using housekeeping genes (expressed across all types) and cell-type-specific marker genes, with Poisson-sampled counts and controlled noise levels. For scalability testing, we varied the number of cell types (50, 100, 200, 500 types), while for robustness analysis, we fixed the dataset at 50 cell types and varied noise levels (1.0, 2.0, 4.0 ,8.0). All synthetic experiments used identity matrices as ground truth, enabling precise accuracy quantification through ARI. Detailed information about the data generation process can be found in Appendix~\ref{synthesis}.

\subsubsection{Retinal Cell Datasets}
We also evaluated our method using three established retinal cell datasets:
\begin{itemize}
    \item Macaque BC and RGC data from peripheral or foveal retina reported by Peng et al.~\cite{pengMolecularClassificationComparative2019a}.
    \item Mouse BC dataset from Shekhar et al.~\cite{shekharCOMPREHENSIVECLASSIFICATIONRETINAL2016a}.
    \item Mouse RGC dataset from Tran et al.~\cite{tranSinglecellProfilesRetinal2019a}.
\end{itemize}

We first validated our method on macaque retinal BC types with known correspondences between peripheral and foveal regions. Specifically, we created a ground-truth test case comprising 12 BC types with exact 1:1 matching across these regions (excluding the fovea-specific "OFFx" type). Subsequently, we extended our analyses to cross-species comparisons: mouse BC (15 types) versus macaque BC (12 types), and mouse RGC (45 types) versus macaque foveal RGC (18 types).

\subsection{Preprocessing and Feature Selection}
We adhered to previously established preprocessing and feature selection protocols~\cite{hahnEvolutionNeuronalCell2023a,shekharCOMPREHENSIVECLASSIFICATIONRETINAL2016a,pengMolecularClassificationComparative2019a,tranSinglecellProfilesRetinal2019a,qiaoFactorizedDiscriminantAnalysis2023c,qiaoDecipheringGeneticCode2024a}. Orthologous genes between mouse and macaque were identified using standard orthology databases, resulting in 10,416 common genes. Cell transcript counts were normalized to the median total transcript count per cell, followed by log-transformation to stabilize variance and mitigate technical variability.

For feature selection, we ranked genes by their SNR (Eq.~\ref{eq:SNR}) and retained genes surpassing an empirical threshold (SNR > 100, yielding a manageable number of highly discriminative genes, typically a few hundred to a thousand in our datasets), resulting in a subset of genes that effectively differentiate cell types. Cell-type centroids were computed based on these selected genes, forming input matrices \(\hat{\mathbf{X}}\) and \(\hat{\mathbf{Y}}\) to the OT algorithm.

\subsection{Cell Type Correspondence}
We formulated cell type correspondence as an OT problem with uniform marginals, \(p_{i} = 1/a\) and \(q_{j} = 1/b\). This choice reflects an agnostic prior about correspondences and prevents dominant cell types from disproportionately influencing the correspondence matrix.

To quantify correspondence strength between cell types between cell types $i,j$ from the final transport matrix $\mathbf{W}^*$, we defined an alignment score $s_{ij}$ as:
\begin{equation}
\label{eq:alignmentScore}
s_{ij} = \frac{1}{2}\left(\frac{W^{*}_{ij}}{\sum_{k=1}^{a}{W^{*}_{kj}}} + \frac{W^{*}_{ij}}{\sum_{l=1}^{b}{W^{*}_{il}}}\right).
\end{equation}
This score $s_{ij}$ represents the average of two probabilities: the probability of cell type $i$ in species one matching type $j$ among all cell types in species two, and vice versa. By averaging both row-wise and column-wise normalized weights, this metric robustly identifies and ranks strong bidirectional evolutionary relationships.

For interpretation of cross-species RGC correspondences, we integrated established morphological and functional annotations from prior comprehensive studies~\cite{qiaoDecipheringGeneticCode2024a,goetzUnifiedClassificationMouse2022a}. This allowed mapping of transcriptomic clusters from the mouse dataset~\cite{tranSinglecellProfilesRetinal2019a} to functionally characterized RGC subtypes, such as alpha, direction-selective, and intrinsically photosensitive RGCs, thereby facilitating biological interpretation of identified macaque RGC counterparts.

\subsubsection{Evaluation Metrics}
We used four key metrics to assess our approach:
\begin{itemize}
\item Adjusted Rand Index (ARI): Measures agreement between predicted correspondences and known ground-truth assignments; higher ARI indicates greater accuracy.
\item Entropy: Quantifies uncertainty or diffuseness of correspondence assignments (Eq.~\ref{eq:entropyFunction}); lower entropy indicates clearer correspondences.
\item Sparsity Score: Proportion of negligible entries (<0.001) in the correspondence matrix, indicating interpretability through sparsity.
\item Running Time: Computational efficiency measured by elapsed processing time.
\end{itemize}

\subsubsection{Parameter Selection}
While OT-MESH involves three main hyperparameters: regularization strength ($\alpha$), MESH learning rate ($\lambda$), and number of MESH iterations ($T$), we developed a parameter selection strategy based on the observation that $\lambda$ and $T$ primarily control the sparsity of the resulting transport plan. Our selection heuristic uses an elbow method to identify parameter combinations that achieve sufficient sparsity without over-regularization. Specifically, for each $\alpha$ value, we visualize the relationship between the entropy and different values of $\lambda$ and $T$ to identify the elbow point where the plot shows a sharp bend, indicating diminishing effects of reducing entropy with adjusting parameters (SupplementaryFigure 1A,B). 

From this set of high-sparsity candidates (one for each $\alpha$), we select the final parameter triplet (including $\alpha$) that minimizes the transport cost, thus balancing sparsity with fidelity to the cost matrix (SupplementaryFigure 1C). This approach provides data-driven method for parameter selection that ensures reproducibility while allowing flexibility to adapt to different dataset characteristics. 

To apply this heuristic efficiently across datasets of varying scales, we utilize adaptive search ranges. The parameter ranges are defined as follows:

\begin{itemize}
    \item For smaller-scale problems (< 100 types per species, e.g., retinal cell type comparisons):
    \begin{itemize}
        \item Regularization parameter \(\alpha\): [0.1, 0.5, 1.0, 5.0, 10.0]
        \item MESH learning rate \(\lambda\): [0.1, 0.5, 1.0, 5.0, 10.0]
        \item Number of MESH iterations \(T\): [2.0, 4.0, 6.0, 8.0, 10.0]
    \end{itemize}

    \item For larger-scale problems (\(\geq\) 100 types per species, e.g., scalability analysis): the search ranges are extended to [0.5, 1.0, 5.0, 10.0, 50.0] for \(\lambda\) and to [8.0, 10.0, 12.0, 14.0, 16.0] for \(T\), as for larger and more complex transport matrices, a more aggressive optimization schedule is required to effectively drive the solution towards a sparse and interpretable state.
\end{itemize}

\subsubsection{Comparative Baselines}
To contextualize our method, we compared it against three baseline approaches:

\begin{itemize}
\item XGBoost classification (reference-based matching): A supervised classification approach commonly used in previous cross-species studies~\cite{pengMolecularClassificationComparative2019a, zhangEvolutionaryDevelopmentalSpecialization2024, wangMolecularCharacterizationSea2024}. An XGBoost classifier is trained on cell types from a designated reference species using orthologous gene expression profiles, and applied to predict cell types in a query species. The correspondence matrix is derived by aggregating the predicted labels of individual query cells to quantify cell-type level relationships.

\item Harmony integration followed by 1-Nearest Neighbor (Harmony+1NN, projection-based matching): Harmony~\cite{korsunskyFastSensitiveAccurate2019} integrates individual-cell transcriptomic profiles from different species into a shared low-dimensional embedding, correcting batch effects. Cell-type correspondences are inferred by assigning each cell to the cell type of its nearest neighbor from the other species in the integrated space. These individual-level assignments are then aggregated to form a correspondence matrix reflecting inter-species cell-type alignments.

\item RefCM: An OT-based method~\cite{galantiAutomatedCellType2024} that maps query cell type clusters to reference clusters through controlled complexity parameters: \texttt{max\_merges} (maximum query clusters mapping to one reference cluster), \texttt{max\_splits} (maximum reference clusters per query cluster), and \texttt{discovery\_threshold} (cost threshold above which clusters are considered novel types rather than mapped). We evaluated two configurations to assess performance under different mapping constraints: RefCM-Strict (\texttt{max\_merges/splits}=1, \texttt{discovery\_threshold}=0.0) enforcing strict 1-to-1 mappings comparable to our expected ground truth, and RefCM-Flexible (\texttt{max\_merges/splits} = 10\% of cell types, \texttt{discovery\_threshold}=0.5) permitting complex many-to-many correspondences to test robustness when the true mapping structure is unknown.

\item Entropy-regularized OT without MESH refinement: To explicitly evaluate the benefit of MESH, we included standard entropy-regularized OT as a baseline, using the same SNR-selected genes for fair comparison.

\end{itemize}

All methods were evaluated using the previously described metrics, allowing rigorous comparison of accuracy, interpretability, and computational efficiency.

\section{Results}
\subsection{Application to Synthetic Datasets}

\subsubsection{Computational Scalability Analysis}
To evaluate the performance of OT-MESH, especially its scalability, and compare it against baseline methods, we conducted systematic benchmarks using synthetic datasets with varying numbers of cell types (50, 100, 200, and 500 types). Each dataset maintained 50 cells per type with known 1:1 correspondences, allowing precise quantification of matching accuracy.

\begin{figure}[htbp]
\centering
\includegraphics[trim=0 350 250 0, clip, width=0.95\linewidth]{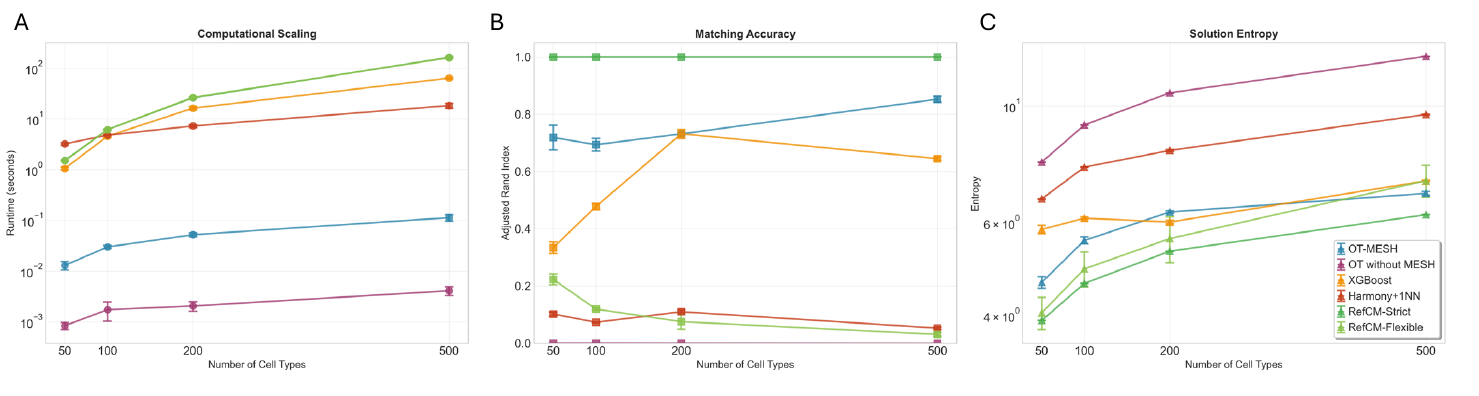}
\caption{Scalability analysis across varying numbers of cell types. A) Runtime scaling of different methods. B) Matching accuracy (ARI) over cell type number of different methods. C) Solution entropy scaling of different methods. Different methods are indicated in the legend of panel C. Error bars represent standard deviation across three independent runs.}
\label{fig2}
\end{figure}

Figure \ref{fig2} presents the scaling behavior across three key metrics. OT-MESH demonstrated favorable computational scaling (Figure \ref{fig2}A), maintaining sub-second runtime across all tested scales (0.013s for 50 types to 0.113s for 500 types). In contrast, XGBoost showed steeper scaling: its runtime increased from 1.052s for 50 types to 60.53s for 500 types due to the growing complexity of multi-class classification. Harmony+1NN exhibited similar challenges, requiring 3.216s for 50 types and 18.266s for 500 types due to the computational cost of embedding and nearest-neighbor search. On average, both RefCM variants showed the highest computational costs among all methods, requiring approximately 160s for the largest dataset due to the complexity of computing optimal transport over earth mover's distances (EMDs).

For matching accuracy (Figure \ref{fig2}B), OT-MESH maintained consistently high performance (ARI > 0.69) across all scales. RefCM-Strict achieved perfect accuracy (ARI = 1.0) as expected, given that it enforces 1-to-1 mappings matching our ground truth structure, though at substantial computational cost. When this constraint was relaxed in RefCM-Flexible, performance degraded significantly, with ARI declining from 0.4 to 0.2 as problem size increased. Both XGBoost and Harmony+1NN underperformed relative to OT-MESH, while standard entropy-regularized OT without MESH completely failed (ARI close to 0) across all scales, confirming the necessity of entropy minimization in our approach.

The entropy analysis (Figure \ref{fig2}C) revealed that OT-MESH solutions maintained low entropy (4.6-6.8) across all scales, indicating concentrated, interpretable correspondence matrices. Standard OT without MESH produced highly diffuse solutions with the highest entropy values, increasing from 7.8 to 12.4 as the number of types grew. RefCM-Strict achieved the lowest entropy values (3.9-6.2), consistent with its enforced 1-to-1 constraint, while other methods showed intermediate values.

\subsubsection{Robustness to Noise}
We evaluated method robustness by systematically varying noise levels (1.0, 2.0, 4.0, and 8.0) on datasets with 50 cell types. Performance degradation patterns across methods are presented in SupplementaryFigure 2.

For matching accuracy (SupplementaryFigure 2A), RefCM-Strict maintained perfect performance (ARI = 1.0) across all noise levels, benefiting from its enforced correspondence structure. OT-MESH demonstrated remarkable robustness with only gradual degradation, maintaining ARI > 0.93 even at the highest noise level. In contrast, XGBoost exhibited steep performance decline, with ARI dropping from 0.67 to 0.05 as noise increased. Harmony+1NN and RefCM-Flexible displayed intermediate robustness, while standard OT without MESH remained ineffective (ARI close to 0) regardless of noise level.

The entropy analysis under varying noise conditions (SupplementaryFigure 2B) revealed distinct response patterns. OT-MESH and RefCM-Strict maintained relatively stable entropy values across all noise levels, indicating that their sparse solutions remained robust to data quality degradation. Conversely, RefCM-Flexible, XGBoost, and Harmony+1NN exhibited increasing entropy with higher noise levels, suggesting their correspondence matrices became progressively more diffuse and uncertain as data quality decreased. Standard OT maintained consistently high entropy across all conditions, reflecting its inherently diffuse nature.

These noise robustness experiments, combined with our scalability analysis, demonstrate that OT-MESH achieves an optimal balance of desirable properties: near-optimal accuracy without requiring prior knowledge of correspondence structure, sub-second runtime even for large-scale problems, and robust performance under challenging noise conditions. While RefCM-Strict guarantees perfect matching when the true correspondence structure is known a priori, OT-MESH provides a practical and efficient solution for real-world scenarios where such knowledge is unavailable.


\subsection{Application to Retinal Cell Datasets}

\subsubsection{Method Validation and Comparison}
Applying to the retinal cell datasets, we first validated OT-MESH against known cell type correspondences between macaque peripheral and foveal bipolar cells (BCs), comparing its performance to baseline methods. For XGBoost, the classifier was trained using macaque foveal BCs as the reference~\cite{pengMolecularClassificationComparative2019a}. For Harmony+1NN, we performed embedding-based integration followed by cell type label transfer through nearest neighbor assignment in the harmonized PCA space~\cite{korsunskyFastSensitiveAccurate2019}. RefCM-Strict was configured to enforce 1-to-1 mappings~\cite{galantiAutomatedCellType2024}. Standard OT was performed using the same parameters as OT-MESH but without iterative entropy minimization.

\begin{figure}[htbp]
\centering
\includegraphics[trim=0 320 40 0, clip, width=0.95\linewidth]{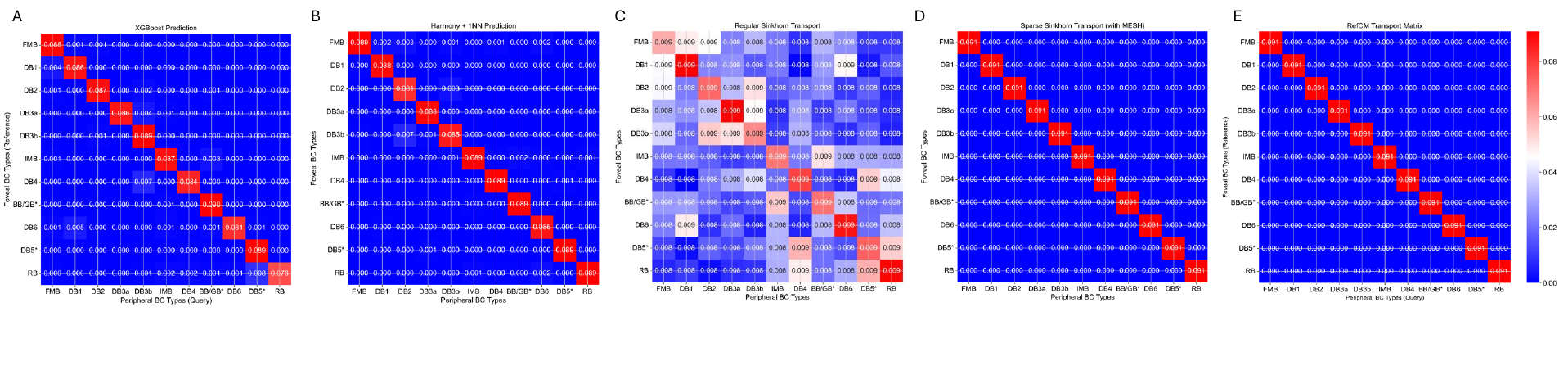}
\caption{Comparison of correspondence matrices for macaque peripheral to foveal BC type mapping. A) XGBoost shows diagonal structure with notable off-diagonal noise. B) Harmony+1NN improves alignment with cleaner structure. C) Standard OT produces a diffuse, uninterpretable mapping. D) OT-MESH yields perfect diagonal structure with high-confidence 1:1 matchings without requiring prior constraints. E) RefCM-Strict achieves clean diagonal through enforced 1-to-1 constraints. Correspondence matrices are normalized so that the sum of all their entries is one.}
\label{fig3}
\end{figure}

Figure \ref{fig3} illustrates the resulting correspondence matrices. XGBoost (Figure \ref{fig3}A) yielded a predominantly diagonal structure with notable off-diagonal noise. Harmony+1NN (Figure \ref{fig3}B) improved upon XGBoost by producing cleaner correspondences, though some off-diagonal elements persisted. Standard OT without MESH refinement (Figure \ref{fig3}C) produced a highly diffuse correspondence matrix, making interpretation impossible despite its computational speed. In contrast, OT-MESH (Figure \ref{fig3}D) achieved a clean diagonal structure with high-confidence values, effectively recovering the true 1:1 correspondences without requiring any prior constraints. RefCM-Strict (Figure \ref{fig3}E) also achieved perfect diagonal alignment through its enforced 1-to-1 mappings.

Quantitative metrics (SupplementaryTable~\ref{table1}) reveal a striking finding: OT-MESH achieved identical perfect performance to RefCM-Strict (sparsity: 0.9091, entropy: 2.3979, ARI: 1.0000) while being 54× faster (0.6375s vs. 34.4387s). This demonstrates that the MESH procedure effectively discovers the optimal sparse correspondence structure through entropy minimization alone, without requiring explicit structural constraints. Harmony+1NN showed good performance (ARI: 0.9258) but required 23× longer runtime than OT-MESH. XGBoost exhibited lower accuracy (ARI: 0.8871) and was 3.6× slower. While standard OT was fastest (0.0296s), its extremely low sparsity, negligible ARI, and diffuse nature rendered it unsuitable for biological interpretation. These results establish that OT-MESH uniquely combines the accuracy of constrained methods with superior computational efficiency, providing an effective solution for cross-species cell type correspondence.


\subsubsection{Cross-Species Correspondence of Bipolar Cell (BC) Types}
We next applied OT-MESH to cross-species alignment of BC types between mouse (15 types) and macaque (12 types). The resulting correspondence matrix (Figure \ref{fig4}A) and matched cell type graph (SupplementaryFigure 3A) reveal clear evolutionary matches, forming strong diagonal-like patterns.

Alignment scores (SupplementaryTable~\ref{table2}) quantify these correspondences. The strongest match was between mouse RBC and macaque RB (rod BCs), with a score close to 1. Among cone BCs, macaque DB1, DB2, DB3a, DB3b, DB6, IMB, and BB/GB* types were uniquely and strongly assigned to mouse BC2, BC4, BC3A, BC3B, BC6, BC7, and BC8/9 types, respectively, consistent with previous reports~\cite{hahnEvolutionNeuronalCell2023a, shekharCOMPREHENSIVECLASSIFICATIONRETINAL2016a}.

The remaining three macaque BC types showed more complex mappings. Macaque FMB corresponded to both mouse BC1A and BC1B. Macaque DB4 mapped to both mouse BC5A and BC5D. Notably, macaque DB5* showed distributed correspondence with mouse BC5B and BC5C. The mapping of DB5* (a cone BC) to mouse BC5B/C (cone BCs) is more consistent with their shared functional (ON pathway) and morphological (axon terminal stratification in middle IPL) properties ~\cite{eulerRetinalBipolarCells2014b, tsukamotoBipolarCellsMacaque2016} compared with a previous report mapping it to mouse RBC (a rod BC)~\cite{shekharCOMPREHENSIVECLASSIFICATIONRETINAL2016a}. This highlights OT-MESH's ability to uncover complex homologies.

\subsubsection{Cross-Species Correspondence of Ganglion Cell (RGC) Types}
\begin{figure} [htbp]
\centering
\includegraphics[trim=0 325 150 0, clip, width=0.95\linewidth]{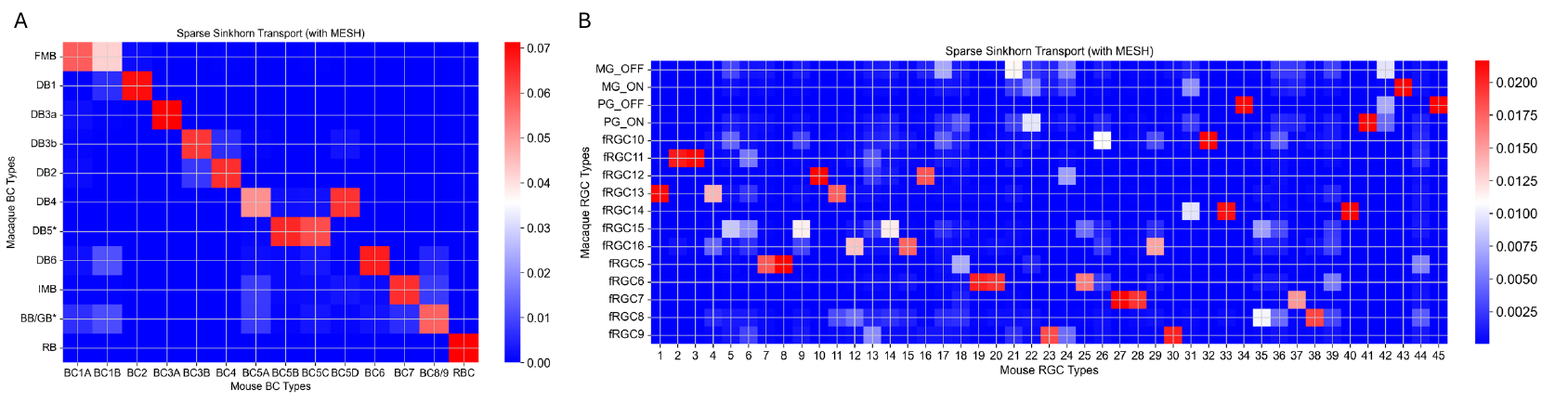}
\caption{Cross-species correspondence matrices. A) Mouse-to-macaque BC correspondence showing clear mapping between evolutionarily conserved types. B) Mouse-to-macaque RGC correspondence revealing more complex relationships due to the difference in type numbers (45 mouse vs. 18 macaque types).}
\label{fig4}
\end{figure}

When applied to the more complex case of RGC types (45 mouse types vs. 18 macaque foveal types), OT-MESH successfully recovered meaningful cross-species correspondences despite the substantial difference in type numbers and greater evolutionary divergence. Figure \ref{fig4}B visualizes these relationships, showing a sparser correspondence matrix compared to BCs, but with several strong matches between functionally similar types, reflecting known conserved pathways and also specialized lineages.

Most notably, our analysis confirmed the correspondence between mouse alpha RGCs and primate midget/parasol ganglion cells with high confidence (SupplementaryTable~\ref{table3}, SupplementaryFigure 3B). Specifically, mouse OFF transient alpha (C45) maps strongly to macaque OFF parasol ganglion cells (PG-OFF) (score: 0.71), and ON sustained alpha (C43) corresponds to macaque ON midget ganglion cells (MG-ON) (score: 0.72). Similarly, mouse ON transient alpha (C41) and OFF sustained alpha (C42) map to macaque ON parasol (PG-ON) and OFF midget ganglion cells (MG-OFF) respectively (scores: 0.67 and 0.30). This provides additional support for the evolutionary relationship between these cell types, consistent with the recent report~\cite{hahnEvolutionNeuronalCell2023a}. The correspondence is particularly compelling as it respects both response polarity (ON vs. OFF) and kinetics (sustained vs. transient), suggesting conservation of fundamental visual processing channels across mammalian evolution.

Importantly, OT-MESH identified a strong correspondence between macaque fRGC12 and mouse type C10 (an ON direction-selective RGC) with a high alignment score of 0.71 (SupplementaryTable~\ref{table3}, SupplementaryFigure 3B). This finding suggests a previously underappreciated evolutionary relationship between these types. Notably, the existence of an ON direction-selective RGC in primate retina was independently confirmed by Wang et al.~\cite{wangONtypeDirectionselectiveGanglion2023}, lending strong external validation to our method's predictive power. We also detected robust correspondences between macaque fRGC14 and mouse melanopsin-expressing ipRGCs (C40 and C33, corresponding to M1 subtypes) (scores: 0.70 and 0.68), confirming the conservation of this non-image-forming visual pathway across mammals; the mapping between macaque fRGC7 and a mouse ON orientation-selective RGC (C27) (score: 0.67), and the mapping between macaque fRGC11 and mouse F-mini-ON cells (C3) (score: 0.67), suggesting potential evolutionary links between these functionally specialized types. The correspondence matrix also revealed that many mouse RGC types lacked clear macaque counterparts, suggesting evolutionary specialization in the murine lineage.

Collectively, these results illustrate that OT-MESH captures both well-known and novel cross-species relationships among RGCs. The increased complexity and divergence in RGC mappings, relative to BCs, align with expectations for output neurons under greater evolutionary selection pressure~\cite{hahnEvolutionNeuronalCell2023a, pengMolecularClassificationComparative2019a}.

\section{Discussion}

\subsection{Summary of Key Findings}
We have presented OT-MESH, an unsupervised computational framework that systematically identifies evolutionary correspondences between cell types across species. The core innovation lies in the integration of the MESH procedure with SNR-based gene selection and centroid-based OT cost matrix construction. This combination transforms standard OT outputs—which are typically diffuse and difficult to interpret—into sparse and biologically meaningful correspondence matrices. Applied to retinal neuron datasets from mouse and macaque, OT-MESH successfully recovered known homologies and revealed novel correspondences, including the prediction of an ON direction-selective RGC in primates that was subsequently validated experimentally~\cite{wangONtypeDirectionselectiveGanglion2023}. The observed patterns of conservation in BCs and greater divergence in RGCs align with current understanding of retinal evolution, while maintaining computational efficiency crucial for practical applications.

\subsection{Methodological Advantages}
OT-MESH demonstrates several key advantages over existing approaches. Unlike reference-based methods such as XGBoost~\cite{pengMolecularClassificationComparative2019a, zhangEvolutionaryDevelopmentalSpecialization2024}, OT-MESH is unsupervised and symmetric, eliminating bias from reference species designation. By operating on cell-type centroids rather than individual cells, it achieves orders of magnitude better computational efficiency than projection-based methods like Harmony~\cite{korsunskyFastSensitiveAccurate2019}. Most critically, the MESH refinement procedure addresses the fundamental limitation of standard entropy-regularized OT by actively promoting sparsity through iterative entropy minimization. This results in correspondence matrices that are both mathematically optimal and biologically interpretable.

Our synthetic experiments demonstrated that OT-MESH achieves performance comparable to RefCM-Strict—which has perfect knowledge of the correspondence structure—while being 54× faster and not requiring any prior constraints. This dramatic speedup stems from fundamental algorithmic differences: RefCM computes exact EMD, which has complexity $O(n^3log(n))$ for n cell types. In contrast, OT-MESH leverages entropy-regularized optimal transport solved via the Sinkhorn algorithm, whose complexity has been proven to be nearly $O(n^2)$. For the large-scale atlases emerging from current consortium efforts—often containing hundreds to thousands of cell types—this complexity difference becomes prohibitive. Combined with OT-MESH's robustness to noise (maintaining ARI > 0.93 even at 8× noise levels), these computational advantages make OT-MESH particularly suitable for systematic comparative analyses across multiple species, organs, or developmental stages.

\subsection{Limitations and Future Directions}

\subsubsection{Dependence on Upstream Analysis}
While our noise robustness experiments demonstrate that OT-MESH maintains high performance even under challenging conditions (ARI > 0.93 at 8× noise), the method still depends on the quality of upstream analyses. High-quality gene orthology mapping is essential for accurate cross-species comparisons, and errors in ortholog assignment could propagate through the analysis. Similarly, the method assumes that input cell types have been appropriately identified through clustering or other means. However, our results suggest that OT-MESH is remarkably robust to technical noise. Future work could explore integration with simultaneous clustering and alignment methods to reduce dependence on pre-defined cell type annotations.

\subsubsection{Uniform Marginals and Unbalanced Transport}
Our current implementation assumes uniform marginal distributions, reflecting an agnostic prior about cell type correspondences. While this works well for the retinal datasets studied here, it may be suboptimal for scenarios with many species-specific cell types or when prior knowledge suggests unequal correspondence probabilities. In cases where one species has evolved numerous specialized cell types absent in the other, the uniform marginal assumption could force spurious matches. Extensions using unbalanced optimal transport~\cite{chizatScalingAlgorithmsUnbalanced2017}, which allows for creation and destruction of mass, could better handle such scenarios by explicitly modeling unmatched types. This would be particularly valuable for comparisons across evolutionarily distant species or for organs undergoing rapid evolutionary change.

\subsubsection{Choice of Features and Distance Metrics}
We chose SNR for gene selection because it specifically identifies genes that distinguish cell types (high between-type variance) from those with high technical noise (high within-type variance), making it well-suited for building robust cell-type centroids. Similarly, cosine distance was selected for its insensitivity to expression magnitude differences that commonly arise from technical or biological variation between species. However, these choices may not be optimal for all datasets. For instance, datasets with strong batch effects might benefit from more sophisticated feature selection methods, while comparisons focusing on absolute expression levels might require alternative distance metrics. Future work could explore adaptive selection of these components based on dataset characteristics, or ensemble approaches that combine multiple feature sets and metrics.

\subsubsection{Extension to Multiple Modalities}
We only tested our methodology on transcriptomic data, but cell type identity is determined by multiple molecular and functional properties. Future versions could incorporate chromatin accessibility, DNA methylation, spatial location, morphology, or electrophysiological features. Multi-modal integration could resolve ambiguities in purely transcriptomic comparisons and provide more comprehensive evolutionary insights. The OT framework naturally extends to such settings through appropriate cost matrix design that combines multiple data types.

\section{Conclusion}
OT-MESH provides a principled, efficient, and interpretable solution to the fundamental problem of cross-species cell type matching. By combining optimal transport theory with entropy minimization, it achieves the accuracy of constrained methods while maintaining the flexibility and efficiency needed for exploratory analysis. As single-cell atlases expand across the tree of life, tools like OT-MESH will be essential for synthesizing these data into a unified understanding of cellular evolution. The framework's extensibility to unbalanced transport, multiple modalities, and hierarchical matching positions it well for future developments in comparative genomics and evolutionary cell biology.

\bibliographystyle{unsrt} 
\bibliography{neurips_2024} 

\begin{thebibliography}{10}

\bibitem{wangTracingCelltypeEvolution2021}
Jingjing Wang, Huiyu Sun, Mengmeng Jiang, Jiaqi Li, Peijing Zhang, Haide Chen, Yuqing Mei, Lijiang Fei, Shujing Lai, Xiaoping Han, Xinhui Song, Suhong Xu, Ming Chen, Hongwei Ouyang, Dan Zhang, Guo-Cheng Yuan, and Guoji Guo.
\newblock Tracing cell-type evolution by cross-species comparison of cell atlases.
\newblock {\em Cell Reports}, 34(9):108803, March 2021.

\bibitem{hahnEvolutionNeuronalCell2023a}
Joshua Hahn, Aboozar Monavarfeshani, Mu~Qiao, Allison~H. Kao, Yvonne K{\"o}lsch, Ayush Kumar, Vincent~P. Kunze, Ashley~M. Rasys, Rose Richardson, Joseph~B. Wekselblatt, Herwig Baier, Robert~J. Lucas, Wei Li, Markus Meister, Joshua~T. Trachtenberg, Wenjun Yan, Yi-Rong Peng, Joshua~R. Sanes, and Karthik Shekhar.
\newblock Evolution of neuronal cell classes and types in the vertebrate retina.
\newblock {\em Nature}, 624(7991):415--424, December 2023.

\bibitem{shaferCrossSpeciesAnalysisSingleCell2019}
Maxwell E.~R. Shafer.
\newblock Cross-{{Species Analysis}} of {{Single-Cell Transcriptomic Data}}.
\newblock {\em Front. Cell Dev. Biol.}, 7, September 2019.

\bibitem{shekharCOMPREHENSIVECLASSIFICATIONRETINAL2016a}
Karthik Shekhar, Sylvain~W. Lapan, Irene~E. Whitney, Nicholas~M. Tran, Evan~Z. Macosko, Monika Kowalczyk, Xian Adiconis, Joshua~Z. Levin, James Nemesh, Melissa Goldman, Steven~A. McCarroll, Constance~L. Cepko, Aviv Regev, and Joshua~R. Sanes.
\newblock {{COMPREHENSIVE CLASSIFICATION OF RETINAL BIPOLAR NEURONS BY SINGLE-CELL TRANSCRIPTOMICS}}.
\newblock {\em Cell}, 166(5):1308--1323.e30, August 2016.

\bibitem{pengMolecularClassificationComparative2019a}
Yi-Rong Peng, Karthik Shekhar, Wenjun Yan, Dustin Herrmann, Anna Sappington, Gregory~S. Bryman, Tav{\'e} {van Zyl}, Michael Tri~H. Do, Aviv Regev, and Joshua~R. Sanes.
\newblock Molecular {{Classification}} and {{Comparative Taxonomics}} of {{Foveal}} and {{Peripheral Cells}} in {{Primate Retina}}.
\newblock {\em Cell}, 176(5):1222--1237.e22, February 2019.

\bibitem{zhangEvolutionaryDevelopmentalSpecialization2024}
Lin Zhang, Martina Cavallini, Junqiang Wang, Ruiqi Xin, Qiangge Zhang, Guoping Feng, Joshua~R. Sanes, and Yi-Rong Peng.
\newblock Evolutionary and developmental specialization of foveal cell types in the marmoset.
\newblock {\em Proc Natl Acad Sci U S A}, 121(16):e2313820121, April 2024.

\bibitem{wangMolecularCharacterizationSea2024}
Junqiang Wang, Lin Zhang, Martina Cavallini, Ali Pahlevan, Junwei Sun, Ala Morshedian, Gordon~L. Fain, Alapakkam~P. Sampath, and Yi-Rong Peng.
\newblock Molecular characterization of the sea lamprey retina illuminates the evolutionary origin of retinal cell types.
\newblock {\em Nat Commun}, 15(1):10761, December 2024.

\bibitem{butlerIntegratingSinglecellTranscriptomic2018b}
Andrew Butler, Paul Hoffman, Peter Smibert, Efthymia Papalexi, and Rahul Satija.
\newblock Integrating single-cell transcriptomic data across different conditions, technologies, and species.
\newblock {\em Nat Biotechnol}, 36(5):411--420, May 2018.

\bibitem{korsunskyFastSensitiveAccurate2019}
Ilya Korsunsky, Nghia Millard, Jean Fan, Kamil Slowikowski, Fan Zhang, Kevin Wei, Yuriy Baglaenko, Michael Brenner, Po-ru Loh, and Soumya Raychaudhuri.
\newblock Fast, sensitive and accurate integration of single-cell data with {{Harmony}}.
\newblock {\em Nat Methods}, 16(12):1289--1296, December 2019.

\bibitem{rosenUniversalCellEmbeddings2024a}
Yanay Rosen, Maria Brbi{\'c}, Yusuf Roohani, Kyle Swanson, Ziang Li, and Jure Leskovec.
\newblock Toward universal cell embeddings: Integrating single-cell {{RNA-seq}} datasets across species with {{SATURN}}.
\newblock {\em Nat Methods}, 21(8):1492--1500, August 2024.

\bibitem{galantiAutomatedCellType2024}
Valerio Galanti, Lingting Shi, Elham Azizi, Yining Liu, and Andrew~J. Blumberg.
\newblock Automated {{Cell Type Annotation}} with {{Reference Cluster Mapping}}, December 2024.

\bibitem{zhangUnlockingSlotAttention2023}
Yan Zhang, David~W. Zhang, Simon {Lacoste-Julien}, Gertjan~J. Burghouts, and Cees G.~M. Snoek.
\newblock Unlocking {{Slot Attention}} by {{Changing Optimal Transport Costs}}, May 2023.
\newblock Comment: Published at International Conference on Machine Learning (ICML) 2023.

\bibitem{tranSinglecellProfilesRetinal2019a}
Nicholas~M. Tran, Karthik Shekhar, Irene~E. Whitney, Anne Jacobi, Inbal Benhar, Guosong Hong, Wenjun Yan, Xian Adiconis, McKinzie~E. Arnold, Jung~Min Lee, Joshua~Z. Levin, Dingchang Lin, Chen Wang, Charles~M. Lieber, Aviv Regev, Zhigang He, and Joshua~R. Sanes.
\newblock Single-cell profiles of retinal neurons differing in resilience to injury reveal neuroprotective genes.
\newblock {\em bioRxiv}, page 711762, July 2019.

\bibitem{wangONtypeDirectionselectiveGanglion2023}
Anna Y.~M. Wang, Manoj~M. Kulkarni, Amanda~J. McLaughlin, Jacqueline Gayet, Benjamin~E. Smith, Max Hauptschein, Cyrus~F. McHugh, Yvette~Y. Yao, and Teresa Puthussery.
\newblock An {{ON-type}} direction-selective ganglion cell in primate retina.
\newblock {\em Nature}, 623(7986):381--386, November 2023.

\bibitem{cuturiSinkhornDistancesLightspeed2013}
Marco Cuturi.
\newblock Sinkhorn {{Distances}}: {{Lightspeed Computation}} of {{Optimal Transportation Distances}}, June 2013.

\bibitem{demetciSCOTSingleCellMultiOmics2022}
Pinar Demetci, Rebecca Santorella, Bj{\"o}rn Sandstede, William~Stafford Noble, and Ritambhara Singh.
\newblock {{SCOT}}: {{Single-Cell Multi-Omics Alignment}} with {{Optimal Transport}}.
\newblock {\em J Comput Biol}, 29(1):3--18, January 2022.

\bibitem{gossiMatchingSingleCells2023}
Federico Gossi, Pushpak Pati, Panagiotis Chouvardas, Adriano~Luca Martinelli, Marianna {Kruithof-de Julio}, and Maria~Anna Rapsomaniki.
\newblock Matching single cells across modalities with contrastive learning and optimal transport.
\newblock {\em Briefings in Bioinformatics}, 24(3):bbad130, May 2023.

\bibitem{penaConstructingCelltypeTaxonomy2025}
Sebastian Pena, Lin Lin, and Jia Li.
\newblock Constructing {{Cell-type Taxonomy}} by {{Optimal Transport}} with {{Relaxed Marginal Constraints}}.
\newblock {\em arXiv:2501.18650}, January 2025.

\bibitem{qiaoFactorizedDiscriminantAnalysis2023c}
Mu~Qiao.
\newblock Factorized discriminant analysis for genetic signatures of neuronal phenotypes.
\newblock {\em Front. Neuroinform.}, 17, December 2023.

\bibitem{qiaoDecipheringGeneticCode2024a}
Mu~Qiao.
\newblock Deciphering the genetic code of neuronal type connectivity through bilinear modeling.
\newblock {\em Elife}, 12:RP91532, June 2024.

\bibitem{goetzUnifiedClassificationMouse2022a}
Jillian Goetz, Zachary~F. Jessen, Anne Jacobi, Adam Mani, Sam Cooler, Devon Greer, Sabah Kadri, Jeremy Segal, Karthik Shekhar, Joshua~R. Sanes, and Gregory~W. Schwartz.
\newblock Unified classification of mouse retinal ganglion cells using function, morphology, and gene expression.
\newblock {\em Cell Rep}, 40(2):111040, July 2022.

\bibitem{eulerRetinalBipolarCells2014b}
Thomas Euler, Silke Haverkamp, Timm Schubert, and Tom Baden.
\newblock Retinal bipolar cells: Elementary building blocks of vision.
\newblock {\em Nature Reviews Neuroscience}, 15(8):507--519, August 2014.

\bibitem{tsukamotoBipolarCellsMacaque2016}
Yoshihiko Tsukamoto and Naoko Omi.
\newblock {{ON Bipolar Cells}} in {{Macaque Retina}}: {{Type-Specific Synaptic Connectivity}} with {{Special Reference}} to {{OFF Counterparts}}.
\newblock {\em Front. Neuroanat.}, 10, October 2016.

\bibitem{chizatScalingAlgorithmsUnbalanced2017}
Lenaic Chizat, Gabriel Peyr{\'e}, Bernhard Schmitzer, and Fran{\c c}ois-Xavier Vialard.
\newblock Scaling {{Algorithms}} for {{Unbalanced Transport Problems}}, May 2017.

\end{thebibliography}

\section{Data and Code Availability}
\label{availability}
All datasets used in this study are publicly available and cited in the manuscript. The code implementing the described OT-MESH framework for evolutionary cell type matching is available at https://github.com/muqiao0626/Evo-Cell-Type-OT-MESH. All experiments used Python 3.10.12 and runtime measurements were done on Apple M1 Max with 64GB RAM.

\section{Appendix}

\subsection{Appendix A: Synthetic Data Generation}
\label{synthesis}

We developed a biologically-inspired synthetic data generation pipeline to enable controlled evaluation of method performance under varying conditions.

\textbf{Cell-type Expression Profile:} For each of $n$ cell types, we generated expression signatures for $m$ genes divided into two categories:
\begin{itemize}
\item Housekeeping genes : Expressed across all cell types with values drawn from $\text{Uniform}(3,7)$.
\item Marker genes: Cell-type-specific expression with primary markers ($\text{Uniform}(10,15)$) and secondary markers ($\text{Uniform}(5,8)$).
\end{itemize}

\textbf{Cell-level Expression:} Individual cell expression profiles were generated through:
\begin{equation}
x_{ijk} \sim \text{Poisson}(\lambda_{jk}) + \mathcal{N}(0, \sigma^2)
\end{equation}
where $x_{ijk}$ is the expression of gene $k$ in cell $i$ of type $j$, $\lambda_{jk}$ is the signature expression level, and $\sigma$ is the noise level parameter.

\textbf{Cross-species Simulation:} To simulate species differences, we applied uniform scaling where each cell type's expression profile in the second species was multiplied by a factor drawn from $\text{Uniform}(0.8,1.2)$, preserving relative expression patterns while introducing overall expression level differences.

\textbf{Parameter Settings:}
\begin{itemize}
\item Scalability experiments: $n \in \{50, 100, 200, 500\}$ cell types, 50 cells per type, 1000 genes, noise level = 1.0
\item Robustness experiments: $n = 50$ cell types, noise level $\sigma \in \{1.0, 2.0, 4.0, 8.0\}$
\item All experiments: 3 independent runs to get metric statistics
\end{itemize}

Ground truth correspondence matrices were identity matrices, representing perfect 1:1 matching between species, allowing exact quantification of matching accuracy.

\subsection{Appendix B: Supplementary Materials}
\label{supplementary}

\setcounter{figure}{0}  
\renewcommand{\thefigure}{}  
\renewcommand{\figurename}{}  
\makeatletter
\renewcommand{\fnum@figure}{}  
\makeatother

\begin{figure}[H]
\centering
\includegraphics[trim=0 250 25 0, clip, width=0.95\linewidth]{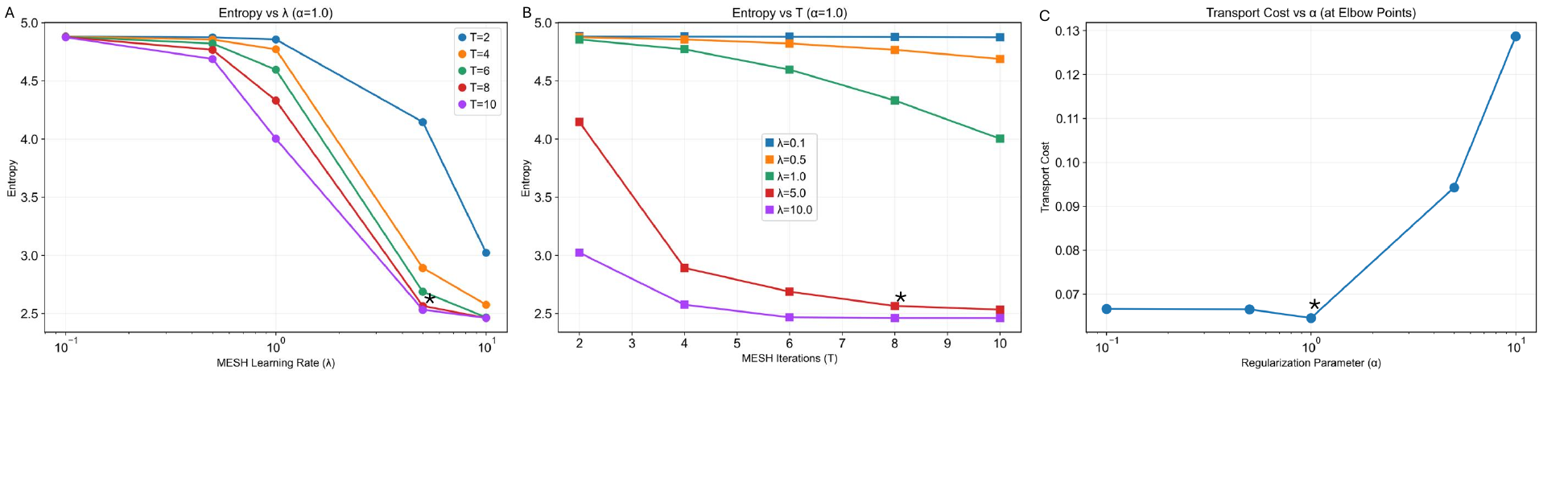}
\caption{Supplementary Figure 1: Parameter selection of OT-MESH, illustrated by the example of the correspondence between macaque peripheral and foveal BC types. A) Entropy versus MESH learning rate ($\lambda$) for $\alpha = 1.0$, showing curves for different numbers of MESH iterations (T). The asterisk marks the elbow point at $\lambda = 5.0$, where further increases in learning rate yield diminishing returns in entropy reduction. B) Entropy versus MESH iterations (T) for $\alpha = 1.0$, showing curves for different learning rates. The asterisk indicates the elbow point at T = 8, beyond which additional iterations provide minimal entropy reduction. C) Transport cost versus regularization parameter ($\alpha$) evaluated at the elbow points identified for each $\alpha$ value. The asterisk marks the optimal parameter combination ($\alpha = 1.0$, $\lambda = 5.0$, T = 8) that minimizes transport cost while maintaining high sparsity. This systematic approach ensures reproducible parameter selection that balances biological interpretability (through sparsity) with fidelity to the underlying gene expression similarities (through transport cost minimization).}
\label{sf1}
\end{figure}

\begin{figure}[H]
\centering
\includegraphics[trim=0 300 300 0, clip, width=0.95\linewidth]{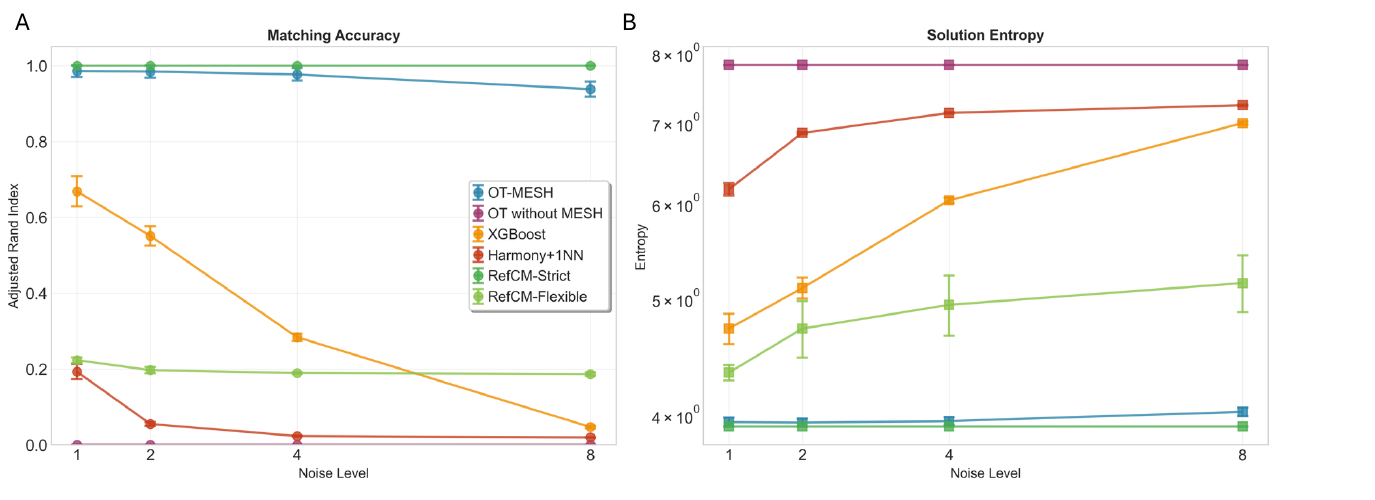}
\caption{Supplementary Figure 2: Robustness analysis across varying noise levels. A) Matching accuracy (ARI) over noise level of different methods. B) Solution entropy over noise level of different methods. Different methods are indicated in the legend of panel A. Error bars represent standard deviation across three independent runs.}
\label{sf2}
\end{figure}

\begin{figure}[H]
\centering
\includegraphics[trim=0 50 450 0, clip, width=0.95\linewidth]{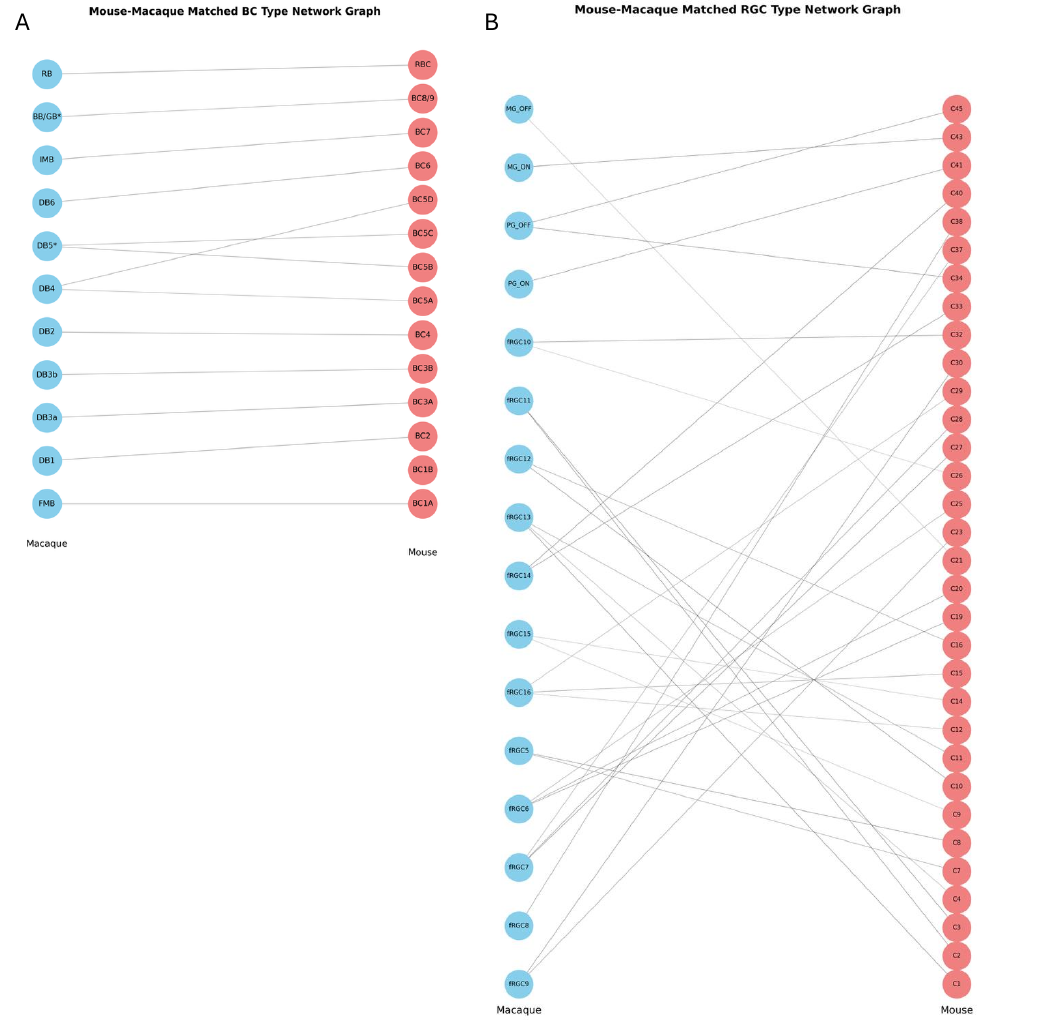}
\caption{Supplementary Figure 3: Cross-species matched cell type network graphs. A) Mouse-Macaque matched BC type network graph. B) Mouse-macaque matched RGC type network graph.}
\label{sf3}
\end{figure}

\setcounter{table}{0}  
\renewcommand{\thetable}{\arabic{table}}  
\renewcommand{\tablename}{Supplementary Table}  

\begin{table}[H]
\centering
\caption{Performance comparison of cross-species cell type matching methods on macaque foveal vs. peripheral BC types. Best values for each metric are in bold.}
\label{table1}
\begin{tabular}{lrrrr}
\toprule
Method          & Sparsity Score & Entropy         & ARI             & Running Time (s) \\ 
\midrule
XGBoost         & 0.8017         & 2.6738          & 0.8871          & 2.3062           \\
Harmony+1NN     & 0.8347         & 2.5949          & 0.9258          & 14.5306          \\
OT without MESH & 0.0000         & 4.7950          & 0.0002          & \textbf{0.0296}  \\ 
OT-MESH         & \textbf{0.9091}& \textbf{2.3979} & \textbf{1.0000} & 0.6375           \\
RefCM-Strict    & \textbf{0.9091}& \textbf{2.3979} & \textbf{1.0000} & 34.4387          \\
\bottomrule
\end{tabular}
\end{table}

\begin{table}[H]
\centering
\caption{Top-ranked correspondences between macaque and mouse BC types quantified by alignment scores (Eq.~\ref{eq:alignmentScore}).}
\label{table2}
\begin{tabular}{llr}
\toprule
Macaque BC Types (Peng et al.\ 2019) & Mouse BC Types (Shekhar et al.\ 2016) & Alignment Score \\ 
\midrule
RB            & RBC   & 0.9979 \\ 
DB3a          & BC3A  & 0.9813 \\ 
DB1           & BC2   & 0.9427 \\ 
DB2           & BC4   & 0.8889 \\ 
DB3b          & BC3B  & 0.8732 \\ 
DB6           & BC6   & 0.8332 \\ 
IMB           & BC7   & 0.8158 \\ 
DB5\textsuperscript{*} & BC5B  & 0.7180 \\ 
DB4           & BC5D  & 0.7162 \\ 
BB/GB\textsuperscript{*} & BC8/9 & 0.7014 \\ 
FMB           & BC1A  & 0.6886 \\ 
DB5\textsuperscript{*} & BC5C  & 0.6605 \\ 
DB4           & BC5A  & 0.5772 \\ 
FMB           & BC1B  & 0.4984 \\ 
\bottomrule
\end{tabular}
\end{table}

\begin{table}[H]
\centering
\caption{Top 10 correspondences between macaque and mouse RGC types quantified by alignment scores. Mouse RGC annotations from Tran et al.~\cite{tranSinglecellProfilesRetinal2019a} and Goetz et al.~\cite{goetzUnifiedClassificationMouse2022a}.}
\label{table3}
\begin{tabular}{lllr}
\toprule
Macaque RGC (Peng et al.\ 2019) & Mouse RGC (Tran et al.\ 2019) & Mouse RGC Annotation & Alignment Score \\ 
\midrule
MG\_ON   & C43 & ON sustained alpha      & 0.7236 \\ 
PG\_OFF  & C45 & OFF transient alpha     & 0.7137 \\ 
fRGC12   & C10 & ON direction-selective  & 0.7079 \\ 
PG\_OFF  & C34 &                         & 0.7064 \\ 
fRGC14   & C40 & M1 ipRGC                & 0.7001 \\ 
fRGC5    & C8  &                         & 0.6938 \\ 
fRGC14   & C33 & M1 ipRGC                & 0.6801 \\ 
fRGC7    & C27 & ON orientation-selective& 0.6740 \\ 
PG\_ON   & C41 & ON transient alpha      & 0.6735 \\ 
fRGC11   & C3  & F-mini-ON               & 0.6714 \\ 
\bottomrule
\end{tabular}
\end{table}

\end{document}